\documentclass[aps,prl,twocolumn,groupedaddress]{revtex4}
\usepackage{graphicx}
\usepackage{dcolumn}
\usepackage{bm}
\usepackage{amssymb}

\begin{document}


\title{Membrane tubule formation by banana-shaped proteins with or without transient network structure} 

\author{Hiroshi Noguchi}
\affiliation{
Institute for Solid State Physics, University of Tokyo,
 Kashiwa, Chiba 277-8581, Japan.
(noguchi@issp.u-tokyo.ac.jp).
}


\begin{abstract}
In living cells,
membrane morphology is regulated by various proteins.
Many membrane reshaping proteins contain a Bin/Amphiphysin/Rvs (BAR) domain, which consists of a banana-shaped rod. 
The BAR domain bends the biomembrane along the rod axis and the features of this anisotropic bending 
have recently been studied.
Here, we report on the role of the BAR protein rods in inducing membrane tubulation, using large-scale coarse-grained simulations. 
We reveal that a small spontaneous side curvature perpendicular to the rod can drastically alter 
the tubulation dynamics at high protein density, whereas no significant difference is obtained at low density. 
A percolated network is intermediately formed depending on the side curvature. 
This network suppresses tubule protrusion, leading to the slow formation of fewer tubules. 
Thus, the side curvature, which is generated by protein--protein and membrane--protein interactions,
plays a significant role in tubulation dynamics.
We also find that positive surface tensions and the vesicle membrane curvature
can stabilize this network structure by suppressing the tubulation. 
\end{abstract}

\maketitle

The Bin/Amphiphysin/Rvs (BAR) superfamily proteins regulate the membrane shape of cell organella as well as membrane fusion and fission; therefore,
BAR protein dysfunction  is implicated in neurodegenerative, cardiovascular, 
and neoplastic diseases \cite{itoh06,masu10,mim12a,zimm06,baum10,joha14,suar14}.
However, the manner in which these proteins assemble on the biomembrane and cooperate to reshape the membranes is not well understood.
The extension of membrane tubes from liposomes
and specific adsorption of the BAR superfamily proteins onto tube regions 
have been observed in {\it in vitro} experiments~\cite{itoh06,masu10,mim12a,matt07,fros08,wang09,sorr12,zhu12,tana13,rame13,shi15}.
Frost et al. have experimentally determined that F-BAR proteins are  adsorbed
on flat regions of lipid membranes using electron microscopy \cite{fros08}. 
Although the assembly seems to constitute the nucleus of the tubule formation,
 the tubule protrusion process has not been experimentally observed.
Recently, Tanaka-Takiguchi et al. reported that the formation dynamics of tubules from a liposome can differ significantly
for different F-BAR proteins \cite{tana13}. That is, FBP17 and CIP4 simultaneously generate many tubule protrusions over the entire liposome surface,
while PSTPIP1 and Pacsin2 generate only a few protrusions from a narrow region of the surface.
In particular, the tubules induced by CIP4 and PSTPIP1 have the same radius.
Thus, the tubule nucleation process depends on the protein type. 
However, it is not known what causes this difference in tubule nucleation behaviour.
Tanaka-Takiguchi et al. also reported that the full length of Pacsin2 induces tubulation, 
but its F-BAR domain region alone does not \cite{tana13}.
In contrast, Wang et al. reported the tubulation is induced by F-BAR domain of Pacsin1 more than by the full-length protein \cite{wang09}.

In the last decade, interactions between laterally isotropic objects on biomembrane, 
such as transmembrane proteins and adsorbed spherical colloids, 
have been intensively investigated \cite{phil09,reyn07,atil07,deme08,auth09,sari12,aimo14}.
In contrast to such studies, however, the interactions between anisotropic adhesives have not yet been explored so far.
The BAR domains are banana shaped and generate an anisotropic curvature
different from the isotropic spontaneous curvature $C_0$ \cite{lipo13}.
This anisotropic nature has recently been receiving increasing theoretical interest.
The classical Canham--Helfrich curvature free energy \cite{canh70,helf73} 
has been extended to anisotropic curvatures \cite{four96,kaba11,igli06}.
Dommersnes and Fournier have derived a many-body potential of long-range interactions 
between point-like anisotropic inclusions and
found linear assemblies and egg-carton membrane structures using Monte Carlo simulations \cite{domm99,domm02}.
In addition, the adsorption and assembly of BAR domains have been investigated using atomic and coarse-grained molecular simulations \cite{arkh08,yu13,simu13,simu15}.
For example, Simunovic et al. have simulated a linear aggregation of N-BAR domains parallel to the domain axis \cite{simu13,simu15}.
However, the relationship between this aggregation and tubulation remains unclear.
Further, tubular formation has been
simulated using a dynamically triangulated membrane model \cite{rama12,rama13} and, also, meshless membrane models \cite{ayto09,nogu14}.
Despite these numerous advancements, 
the present understanding of the physics of membrane shape deformation due to anisotropic curvature is still far from complete.

In this paper, we focus on the effects of the spontaneous (side) curvature $C_{\rm {side}}$ of a protein rod perpendicular to its longest axis 
on the assembly behaviour.
The side curvature has not been focused upon in previous studies,
but here we reveal that it strikingly changes the assembly dynamics.
The excluded volume or van der Waals attraction between proteins and the membrane can effectively generate positive or negative $C_{\rm {side}}$.
We simulate flat membranes and vesicles 
using an implicit-solvent meshless membrane model \cite{nogu09,nogu06,shib11,nogu14,nogu15b},
which allows a large-scale simulation.
A BAR domain is modelled as a banana-shaped rod,
which is assumed to be strongly adsorbed onto the membrane.
The rod length corresponds to $r_{\rm {rod}} \simeq 20$ nm
(the BAR domain lengths range from $13$ to $27$ nm \cite{masu10}).
To investigate the membrane-curvature-mediated interactions,
no direct attractive interaction is considered between the rods.
Our previous studies showed that parallel and perpendicular assemblies
occur separately through membrane-mediated attractive interactions at low protein density \cite{nogu14},
and that polyhedral shapes are formed at high protein density \cite{nogu15b} for vesicles and membrane tubes.

\begin{figure*}
\begin{minipage}[b]{0.4\hsize}
\includegraphics[width=5.cm]{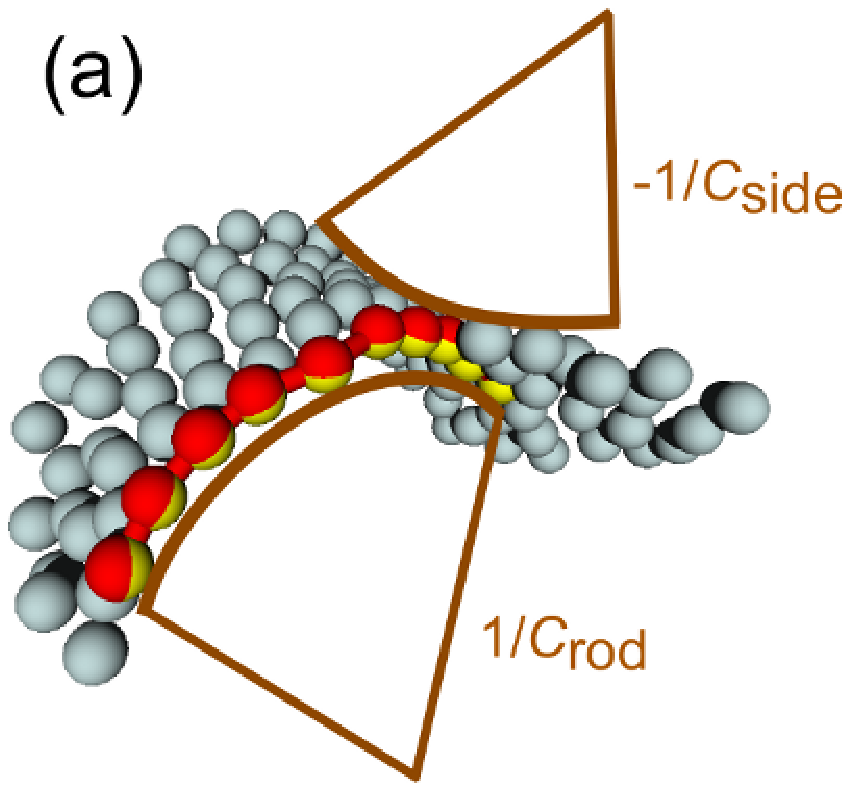}
\includegraphics[width=6.cm]{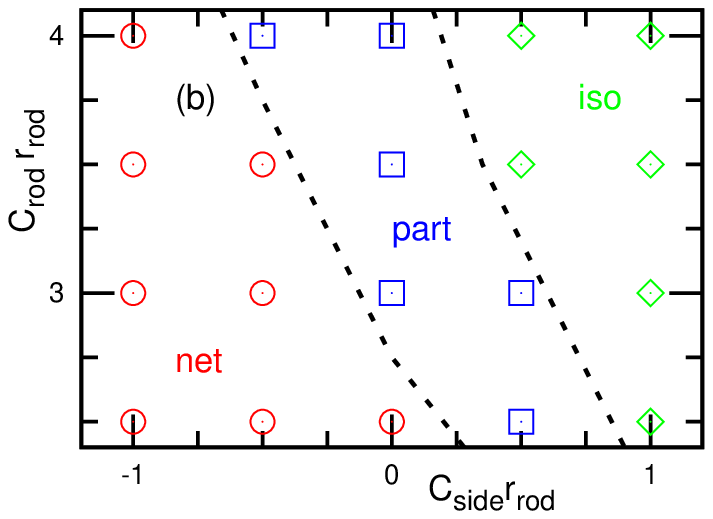}
\end{minipage}
\includegraphics[width=10.cm]{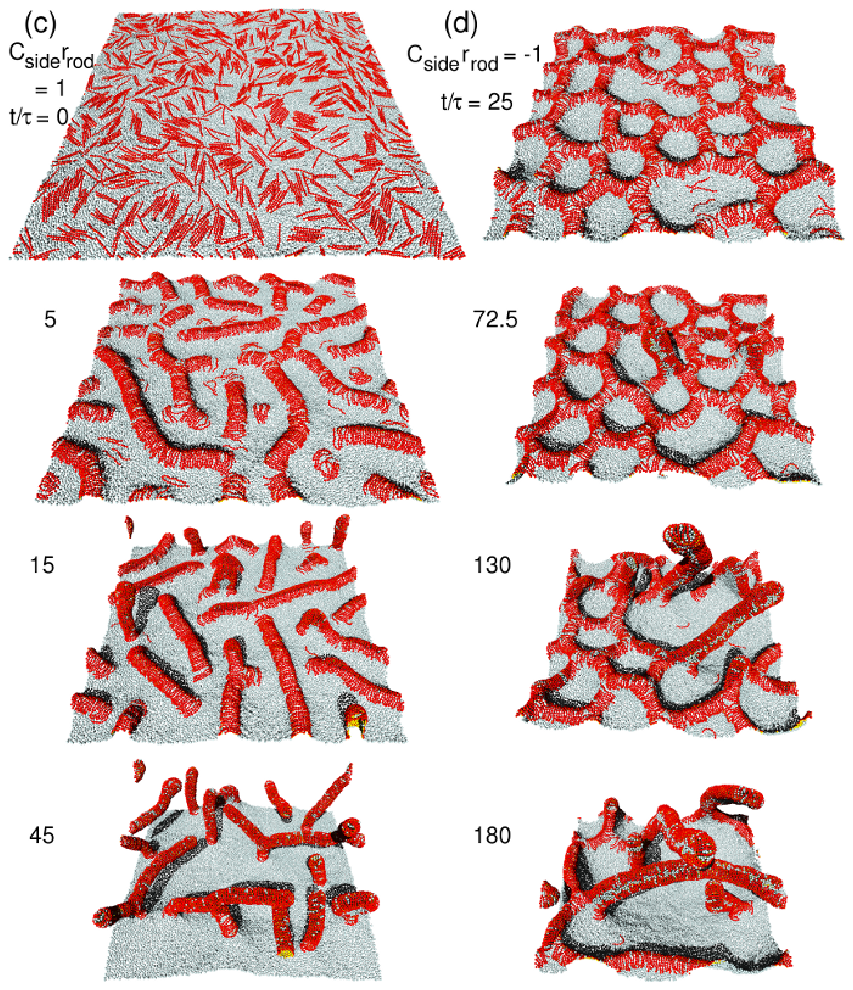}
\includegraphics[width=16cm]{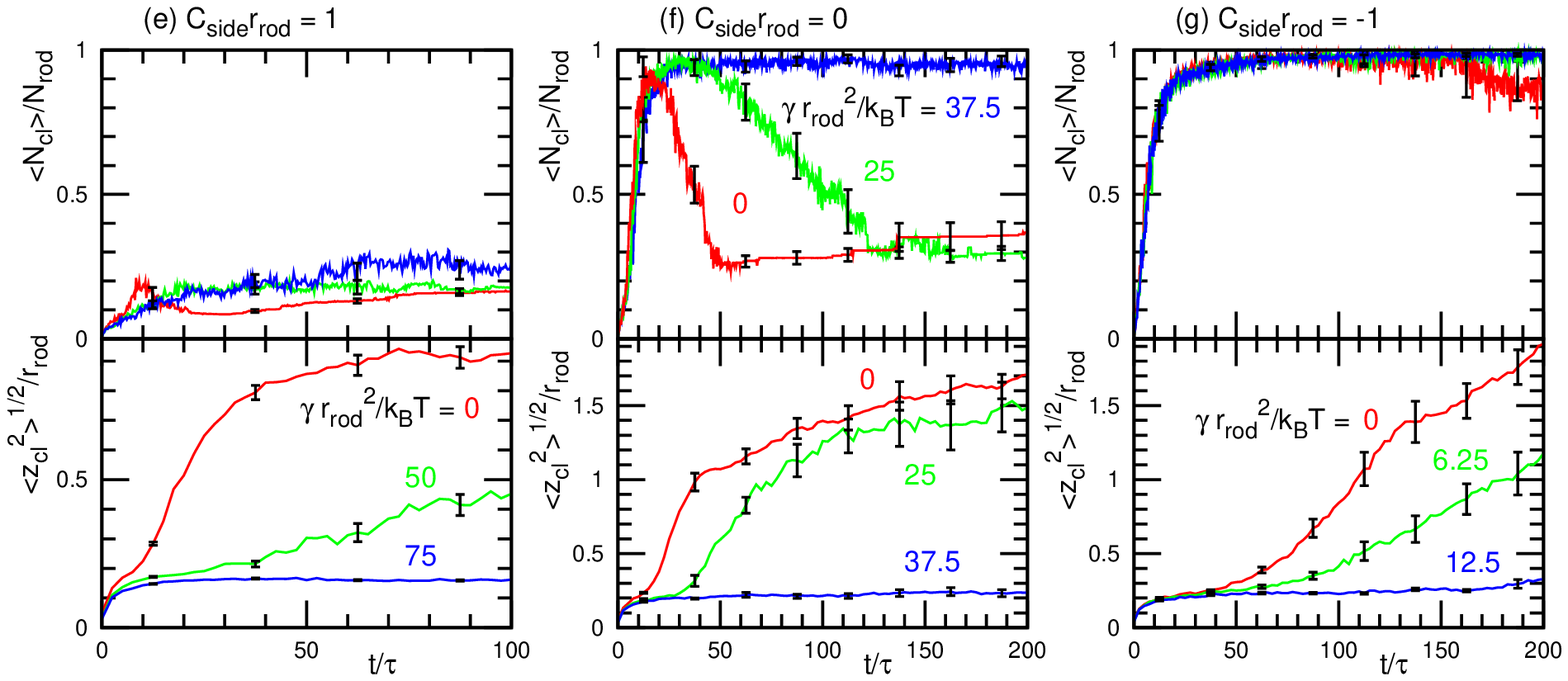}
\caption{
Tubulation dynamics from flat membrane for high rod density, $\phi_{\rm {rod}}=0.4$ ($N_{\rm {rod}}=1,024$).
(a) Protein rod with spontaneous rod and side curvatures, $C_{\rm {rod}}$ and $C_{\rm {side}}$, respectively.
The protein rod is displayed as
a chain of spheres, the halves of which are coloured
red and yellow.
The orientation vector lies along the line of the yellow to red hemispheres.
The light blue spheres represent membrane particles.
(b) Dynamic phase diagram of tubulation from a tensionless flat membrane. 
The red circles represent percolated network formation before tubulation.
The green diamonds indicate that the tubules are formed from isolated clusters.
The blue squares represent partial network formation, and the
dashed lines are guides for the eye.
(c,d) Sequential snapshots of tubulation from tensionless flat membrane
for $C_{\rm {side}}r_{\rm {rod}}=$ (c) $1$ and (d) $-1$ at $C_{\rm {rod}}r_{\rm {rod}}=4$.
(e--g) Time evolution of mean cluster size $\langle N_{\rm {cl}} \rangle$ 
and mean cluster height  $(\langle z_{\rm {cl}}^2 \rangle)^{1/2}$ 
for  $C_{\rm {side}}r_{\rm {rod}}=$ (e) $1$, (f) $0$, 
and (g) $-1$ at  $C_{\rm {rod}}r_{\rm {rod}}=4$.
Error bars calculated from eight independent runs 
are displayed at several data points.
}
\label{fig:snap_n1024}
\end{figure*}

\begin{figure}[t]
\includegraphics[width=6.cm]{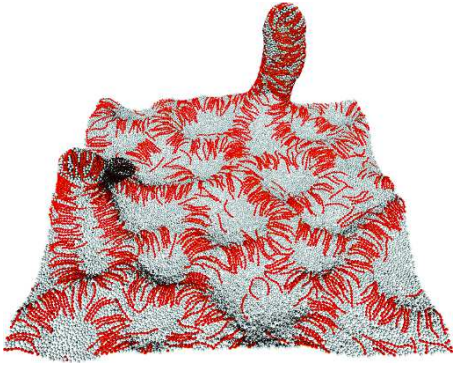}
\caption{
Snapshot of tubulation from tensionless flat membrane 
at $C_{\rm {rod}}r_{\rm {rod}}=2.5$, $C_{\rm {side}}r_{\rm {rod}}=-1$, and $\phi_{\rm {rod}}=0.4$.
}
\label{fig:tube}
\end{figure}

\begin{figure}
\includegraphics{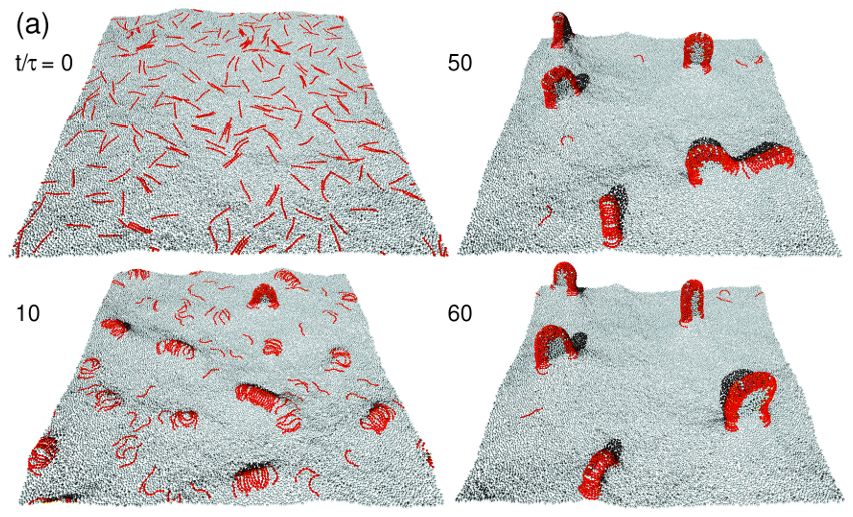}
\includegraphics{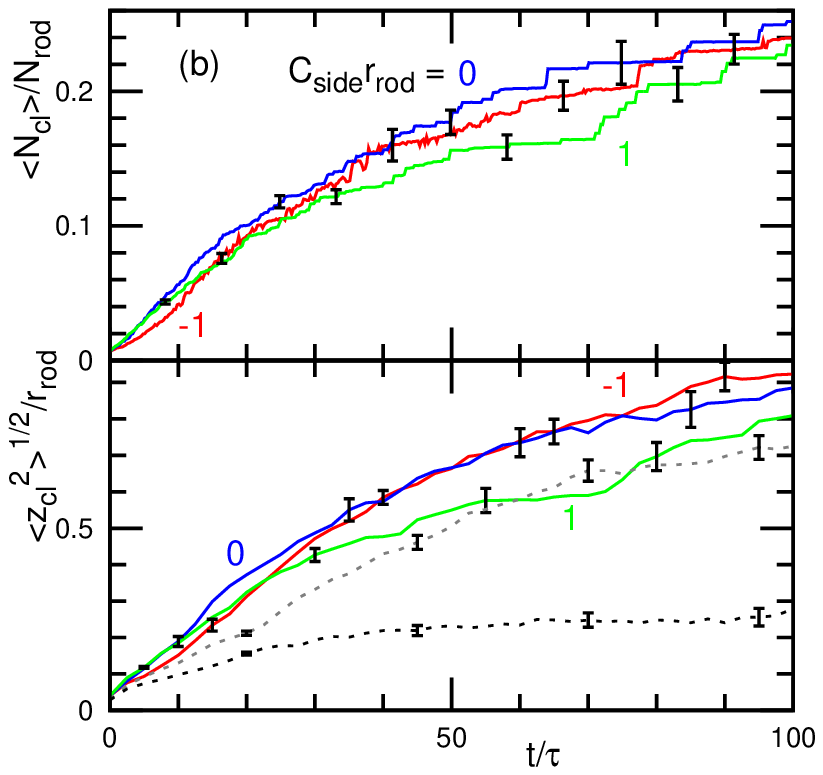}
\caption{
\label{fig:snap_n256}
Membrane tubulation from flat membrane for low $\phi_{\rm {rod}}=0.1$ ($N_{\rm {rod}}=256$), at $C_{\rm {rod}}r_{\rm {rod}}=4$.
(a) Sequential snapshots at $C_{\rm {side}}=0$ and $\gamma=0$.
(b) Time evolution of $\langle N_{\rm {cl}} \rangle$ and $(\langle z_{\rm {cl}}^2 \rangle)^{1/2}$.
The solid lines represent the data for  $C_{\rm {side}}r_{\rm {rod}}=-1$, $0$, and $1$ at $\gamma=0$.
The gray and black  dashed lines represent the data for $\gamma r_{\rm {rod}}^2/k_{\rm B}T=25$ and $37.5$ at $C_{\rm {side}}=0$,
respectively.
Error bars calculated from eight independent runs 
are displayed at several data points.
}
\end{figure}

\section*{Results}

\subsection*{Tubulation from Flat Membrane}

First, we investigate the tubulation from a tensionless flat membrane 
(surface tension $\gamma=0$)
at a high rod density, $\phi_{\rm {rod}}=0.4$
(see Fig.~\ref{fig:snap_n1024}).
The protein rods are initially equilibrated with the rod curvature $C_{\rm {rod}}=0$ and $C_{\rm {side}}=0$.
Once the spontaneous curvatures are altered at $t=0$,
 the rods begin to assemble perpendicularly to the rod axis.
For a positive spontaneous curvature of $C_{\rm {side}}r_{\rm {rod}}=1$ ($C_{\rm {side}}C_{\rm {rod}}>0$),
many tubules simultaneously protrude via the bending of straight rod assemblies
(see Fig.~\ref{fig:snap_n1024}(c) and Supplemental Movie~1). 
Branches of the rod network are formed on the membrane for a short time only.
When the tubulation is initiated, neighbouring branches are broken through lateral shrinkage of the rod assembly.

For a negative  curvature  $C_{\rm {side}}r_{\rm {rod}}=-1$,
the rods form a percolated network covering the entire membrane area [see the top snapshot in Fig.~\ref{fig:snap_n1024}(d)]
and  a tubule protrudes under membrane undulation (see the second snapshot in Fig.~\ref{fig:snap_n1024}(d) and Supplemental Movie~2).
Subsequently, tubule growth occurs along the network.
Thus, the tubulation dynamics is altered remarkably by a relatively small $C_{\rm {side}}$.
Negative and positive $C_{\rm {side}}$ values stabilize and destabilize the network branches, respectively.
The tubulation at $C_{\rm {side}}r_{\rm {rod}}=-1$ is significantly slower than that at  $C_{\rm {side}}r_{\rm {rod}}=1$
and much fewer tubules protrude: the average protrusion time of the first tubule are 
$\langle t_{\rm {tb}}\rangle/\tau=69\pm 4$ and $10\pm 1$
for $C_{\rm {side}}r_{\rm {rod}}=-1$ and $1$ at $C_{\rm {rod}}r_{\rm {rod}}=4$, respectively.

Such characteristic dynamics is distinguishable from the time evolution of the mean cluster size $\langle N_{\rm {cl}}\rangle$
and the root mean square cluster height $(\langle z_{\rm {cl}}^2\rangle)^{1/2}$, 
as shown in Figs.~\ref{fig:snap_n1024}(e)--(g).
For $C_{\rm {side}}r_{\rm {rod}}=-1$, the majority of the rods belong to one large percolated cluster during the tubulation.
In contrast, for $C_{\rm {side}}r_{\rm {rod}}=1$, $\langle N_{\rm {cl}}\rangle$ decreases as the tubules are formed 
and the rod assemblies are divided;
subsequently, $\langle N_{\rm {cl}}\rangle$ slowly increases owing to tubule fusion.
Branched tubules are formed by this fusion [see the bottom snapshot in Fig.~\ref{fig:snap_n1024}(c) and the late stage of Supplemental Movie~1].
Based on the evolution of $\langle N_{\rm {cl}}\rangle$, the tubulation pathways are categorized into three groups [see Fig.~\ref{fig:snap_n1024}(b)]:
tubulation via percolated-network formation (net), via partial-percolated-network formation (part), and without percolation (iso).
When a percolated network does not cover the entire membrane surface or a large cluster of $N_{\rm {cl}} \simeq N_{\rm {rod}}$ 
is maintained for a period shorter than $20\tau$, we categorize the tubulation pathway as part.
 (A typical dynamics is shown in Supplemental Movie~3). 
For the entire parameter range explored in Fig.~\ref{fig:snap_n1024}(b), the final structures are tubules. 
As $C_{\rm {rod}}$ decreases, the tubulation decelerates and a smaller number of large tubules are formed.
The tubules are nucleated and grow from the network vertices at $C_{\rm {rod}}r_{\rm {rod}}=2.5$ or $0.3$ and $C_{\rm {side}}r_{\rm {rod}}=-1$
(see Fig.~\ref{fig:tube}).
The tubule radius $R_{\rm {tb}}$ is roughly determined by $C_{\rm {rod}}$ as $R_{\rm {tb}} \sim 1/C_{\rm {rod}}$.
At $C_{\rm {rod}}r_{\rm {rod}}=4$, the tubule with circumference $2\pi R_{\rm {tb}}\simeq 2r_{\rm {rod}}$ consists of two hemicylinders of the rod assembly.

Our simulation results show that the network formation suppresses the tubulation.
To confirm this more clearly, 
the effects of  $C_{\rm {side}}$ on the rod--membrane interaction are investigated.
A percolated network is not formed during the tubulation at a low $\phi_{\rm {rod}}$ of $0.1$.
Rather, the rods assemble into linear clusters and, subsequently, the large clusters ($N_{\rm {cl}}\gtrsim 40$) transform into tubules 
(see Fig.~\ref{fig:snap_n256} and Supplemental Movie~4). 
Although the initial cluster formation is slightly slower for negative curvature, i.e., $C_{\rm {side}}r_{\rm {rod}}=-1$,
no qualitative difference is detected in the tubulation dynamics [see Fig.~\ref{fig:snap_n256}(b)]. 
Thus, we conclude that the suppression of the tubulation at high rod density is caused  by the mesoscale network formation.

Recent experiments have demonstrated that positive surface tension can suppress tubulation by BAR proteins \cite{shi15} and budding by clathrin coats \cite{sale15}.
In our simulation, the positive tension and network formation  cooperate to suppress tubulation 
[see Figs.~\ref{fig:snap_n1024}(e)--(g) and \ref{fig:snap_n256}(b)].
At $\phi_{\rm {rod}}=0.4$,  the critical tension decreases with increasing $C_{\rm {side}}$:
$\gamma r_{\rm {rod}}^2/k_{\rm B}T\simeq 10$ and $70$ for $C_{\rm {side}}r_{\rm {rod}}=-1$ and $1$, respectively,
where $k_{\rm B}T$ denotes the thermal energy.
These are experimentally measurable magnitudes ($\gamma \simeq 0.1$ mJ/m$^2$ and $0.7$ mJ/m$^2$, respectively).
 The assembly of rods into a clustered network is not suppressed by the applied tension.
In contrast, network structure breaking does not occur at higher tensions [see Fig.~\ref{fig:snap_n1024}(f)].
Thus, the network formation is stabilized by the positive tensions.

When a positive tension $\gamma$ is imposed for the coexisting states of network 
and tubules as shown in the second snapshot of Fig.~\ref{fig:snap_n1024}(d),
the tubules continue to grow at $\gamma r_{\rm {rod}}^2/k_{\rm B}T=40$.
However, the tubules shrink at $\gamma r_{\rm {rod}}^2/k_{\rm B}T=70$.
Thus, at the critical tension $\gamma r_{\rm {rod}}^2/k_{\rm B}T \simeq 50$,
 the tubule elongation force $f_{\rm {tb}}$ by the rod assembly is balanced with 
the expansion of the projected membrane area by the surface tension as $f_{\rm {tb}}=2\pi R_{\rm {tb}}\gamma$.
This tension is higher than that required 
to suppress the tubule protrusion from the flat membranes 
($\gamma r_{\rm {rod}}^2/k_{\rm B}T \simeq 30$), because a nucleation barrier exists for the protrusion.

\begin{figure}
\includegraphics[width=7.8cm]{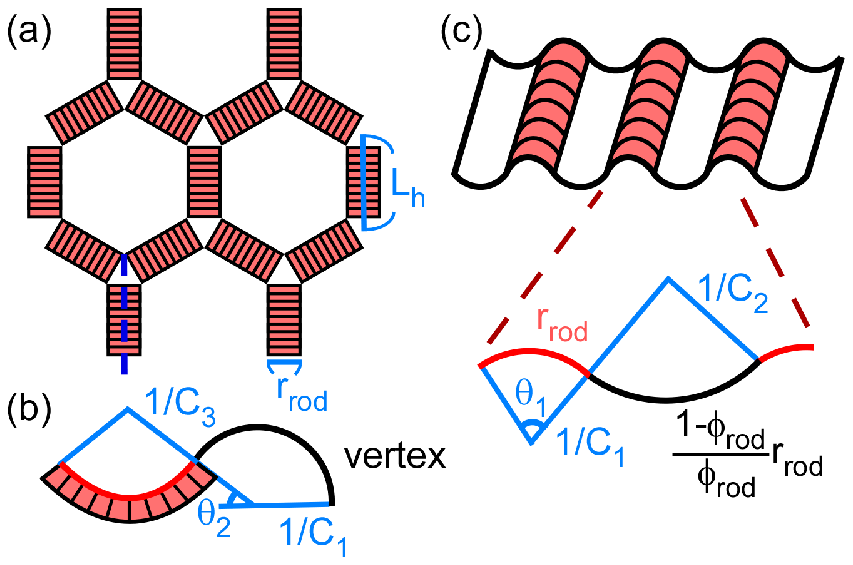}
\includegraphics{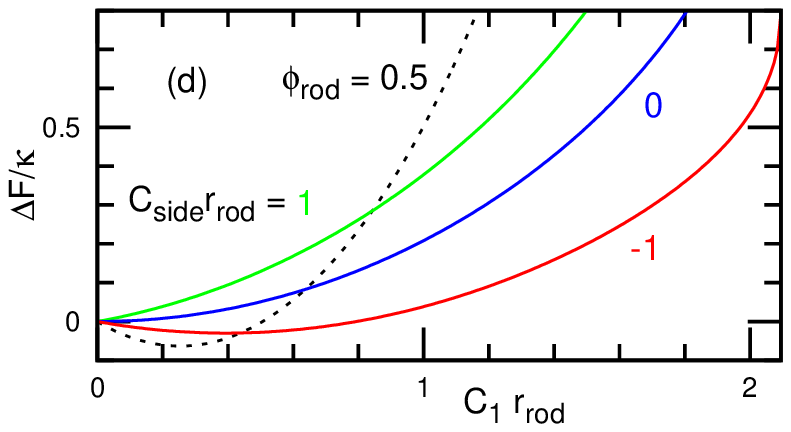}
\caption{
Energy analysis of network structure using simple geometric model.
(a--c) Schematic representation of geometric model.
(a) Top view of hexagonal array at $C_1=0$.
(b) Side view of hexagonal array along the dashed line in (a).
(c) Bird's-eye and front views of striped array.
(d) Free-energy difference $\Delta F$ between hexagonal and striped arrays of rod assembly.
The solid lines represent the data for $C_{\rm {side}}r_{\rm {rod}}=-1$, $0$, and $1$ at $\phi_{\rm {rod}}=0.4$.
The dashed line represents the data for $C_{\rm {side}}r_{\rm {rod}}=-1$ and  $\phi_{\rm {rod}}=0.5$.
}
\label{fig:sch}
\end{figure}

\subsection*{Geometrical Analysis}

To clarify the effects of $C_{\rm {side}}$ on the network formation,
the difference between the free energy of a hexagonal array of the rod assembly and that of a striped array
is estimated using a simple geometric model (see Fig.~\ref{fig:sch}). 
Here, a percolated network is modelled as a hexagonal array with side length $L_{\rm h}$ [Fig.~\ref{fig:sch}(a)],
while an unbranched rod assembly is modelled as a striped array [Fig.~\ref{fig:sch}(c)].
The rod assemblies have rectangular shapes with widths equal to the rod length $r_{\rm {rod}}$.
Our analysis shows that
 the hexagonal network can have lower energy for $C_{\rm {side}}<0$.
This explains why the membranes are trapped in the branched network as a local free-energy minimum.

In the striped array, the rod assemblies are aligned in parallel with intervals
of $(1-\phi_{\rm {rod}})r_{\rm {rod}}/\phi_{\rm {rod}}$.
The rod assemblies are curved upwards and the other regions are curved downwards, 
as shown in Fig.~\ref{fig:sch}(c).
To maintain the continuity of the normal vector of the membrane,
the curvatures have the relation $C_2= \phi_{\rm {rod}}C_1/(1-\phi_{\rm {rod}})$.
The curvature energy $F_{\rm {st}}$ of the striped array per the area $r_{\rm {rod}}^2$ is given by
\begin{eqnarray}
\label{eq:str}
\frac{F_{\rm {st}}}{r_{\rm {rod}}^2} =& & \frac{\kappa_{\rm {r1}}}{2}(C_1-C_{\rm {rod}})^2 \phi_{\rm {rod}}  \\ \nonumber
&+& \frac{\kappa_{\rm {r2}}}{2}C_{\rm {side}}^2 \phi_{\rm {rod}}  + \frac{\kappa}{2} C_2^2 (1-\phi_{\rm {rod}}),
\end{eqnarray}
where $\kappa_{\rm {r1}}$ and $\kappa_{\rm {r2}}$ are the bending rigidities of the rod assembly parallel and perpendicular to the rod axis, respectively. In our simulation, $\kappa=15k_{\rm B}T$,  $\kappa_{\rm {r1}}=40k_{\rm B}T$, and $\kappa_{\rm {r2}}=\kappa$ \cite{nogu15b}.

In the hexagonal array case, the membrane surface is divided into the following three regions: 
Region I: A rectangular rod assembly with length $L_{\rm h}-\sqrt{3}r_{\rm {rod}}/3$ and width $r_{\rm {rod}}$;
Region II: A triangular membrane with  side length $r_{\rm {rod}}$ at the vertex; and
Region III: A hexagonal membrane with  side length $L_{\rm h}-\sqrt{3}r_{\rm {rod}}/3$.
On a flat membrane, the areas of these regions are given by $A_{\rm {I}}= (L_{\rm h}-\sqrt{3}r_{\rm {rod}}/3)r_{\rm {rod}}$,
 $A_{\rm {II}}= \sqrt{3}r_{\rm {rod}}^2/4$, and $A_{\rm {III}}= 3\sqrt{3}(L_{\rm h}-\sqrt{3}r_{\rm {rod}}/3)^2/2$, respectively.
To simplify the calculation, it is assumed that each region has constant curvatures and its area is independent of these curvatures.
Region I has curvatures $C_1$ along the rod axis and $-C_3$ perpendicular to the rod axis
[along the dashed line in Fig.~\ref{fig:sch}(a)], such that it has a saddle shape.
Regions II and III are triangular and hexagonal spherical caps with radii $1/C_1$ and $1/C_4$, respectively.
To maintain the continuity of the normal vector of the membrane,
$C_3= (\pi- 2\theta_2)/(L_{\rm h}-\sqrt{3}r_{\rm {rod}}/3)$ and $C_4= C_1 r_{\rm {rod}}/(\sqrt{3}L_{\rm h}-r_{\rm {rod}})$.
The angle $\theta_2$ is given by $\cos^2(\theta_2)= (1/\cos^2(\theta_1/2)-1)/3$, where $\theta_1= C_1r_{\rm {rod}}$,
since the vertices of three rod assemblies make contact with each other on the spherical cap and maintain three-fold rotational symmetry.
At $\theta_2=0$, $C_1$ has a maximum value of $2\pi/3r_{\rm {rod}}$.
As the area fraction of region I is $\phi_{\rm {rod}}=2\sqrt{3}A_{\rm {I}}/3L_{\rm h}^2$,
 $L_{\rm h}$ is obtained as
\begin{equation}
L_{\rm h} = \frac{2r_{\rm {rod}}}{\sqrt{3}(1 - \sqrt{1-2\phi_{\rm {rod}}})}.
\label{eq:L}
\end{equation}
The curvature energy $F_{\rm {hex}}$ of the hexagonal array per $r_{\rm {rod}}^2$ is given by
\begin{eqnarray}
\label{eq:hex}
\frac{F_{\rm {hex}}}{r_{\rm {rod}}^2} =& & \frac{\kappa_{\rm {r1}}}{2}(C_1-C_{\rm {rod}})^2 \phi_{\rm {rod}}  \\ \nonumber
&+& \frac{\kappa_{\rm {r2}}}{2}(C_3+C_{\rm {side}})^2 \phi_{\rm {rod}} \\ \nonumber
&+& \kappa C_1^2 (1-\phi_{\rm {rod}} - \sqrt{1-2\phi_{\rm {rod}}}) \\ \nonumber
&+& \kappa C_4^2 (1-\phi_{\rm {rod}} + \sqrt{1-2\phi_{\rm {rod}}}).
\end{eqnarray}

The energy difference $\Delta F= F_{\rm {hex}}-F_{\rm {st}}$ is shown in Fig.~\ref{fig:sch}(d).
The first terms in equations~(\ref{eq:str}) and (\ref{eq:hex})  cancel
since both arrays have the same curvature $C_1$ along the rods.
Hence, 
$\Delta F$ is independent of $C_{\rm {rod}}$ and $\kappa_{\rm {r1}}$.
In the hexagonal array, the curvature energy of the rod assembly is reduced for $C_{\rm {side}}<0$,
and $F_{\rm {hex}}$ is smaller than $F_{\rm {st}}$ for small $C_1$.
Thus, the branched network can be stabilized by a negative $C_{\rm {side}}$ during the formation of the rod assembly.

In our simulation, networks with a wider range of $C_{\rm {side}}$ are formed for smaller values of $C_{\rm {rod}}$ [see Fig.~\ref{fig:snap_n1024}(b)].
This dependence can be explained by an effective increase in the area fraction $\phi_{\rm {rod}}$.
The rod assembly region 
contains more  membrane particles with decreasing $C_{\rm {rod}}$ 
(see light blue particles in rod network and tubules in Fig. \ref{fig:tube}).
Thus, the area of region I increases.
At $\phi_{\rm {rod}}=0.5$, the minimum of $\Delta F$ is twice that at $\phi_{\rm {rod}}=0.4$,
although the area fraction is only $25$\% larger [see the dashed line in Fig.~\ref{fig:sch}(d)].
Thus, the effective increase in the rod region enhances the network formation.

As $\phi_{\rm {rod}}$ decreases,  $L_{\rm h}$ of the hexagonal array increases [see equation~(\ref{eq:L})].
However, it is difficult for the hexagonal network with long $L_{\rm h}$ to form spontaneously,
since the tubule formation begins before the rod assembly reaches $L_{\rm h}$.
The simulation results for $\phi_{\rm {rod}}=0.1$ indicate tubulation from clusters with $N_{\rm {cl}}\gtrsim 40$ at $C_{\rm {rod}}r_{\rm {rod}}=4$.
Thus, the formation threshold of the percolated network is $L_{\rm h} \simeq 5r_{\rm {rod}}$,
where $\phi_{\rm {rod}} \simeq 0.2$.
In the simulation at $\phi_{\rm {rod}}=0.2$, the clusters are typically percolated only in one direction 
only at $C_{\rm {rod}}r_{\rm {rod}}=4$ and $C_{\rm {side}}r_{\rm {rod}}=-1$,
which supports this estimation of the critical density.

\begin{figure}
\includegraphics{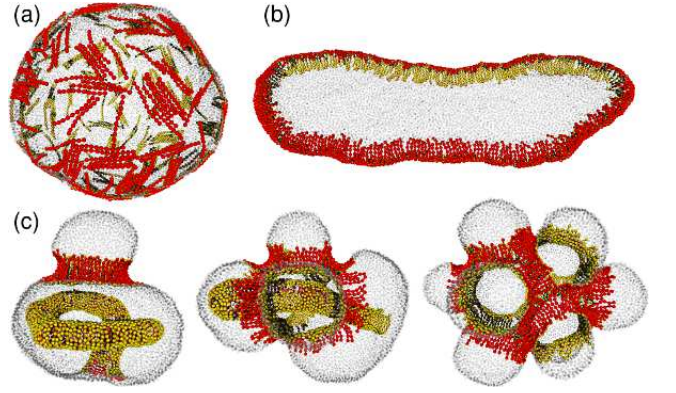}
\caption{ 
\label{fig:snap_sph}
Snapshots of vesicles at $\phi_{\rm {rod}}=0.3$ and $N_{\rm {rod}}=288$.
(a)  Spherical shape in thermal equilibrium  at $C_{\rm {rod}}=0$ and $C_{\rm {side}}=0$.
(b)  Elongated discoidal shape in thermal equilibrium at $C_{\rm {rod}}r_{\rm {rod}}=4$ and $C_{\rm {side}}r_{\rm {rod}}=-1$.
(c) Three metastable shapes at $C_{\rm {rod}}r_{\rm {rod}}=-4$ and $C_{\rm {side}}r_{\rm {rod}}=1$.
The membrane particles are displayed as small transparent spheres for clarity.
}
\end{figure}

\subsection*{Shape Transformation of  Small Vesicles}

Next, we investigate the tubulation from a vesicle of radius $R_{\rm {ves}}=3.07r_{\rm {rod}}$ at $\phi_{\rm {rod}}=0.3$ (see Fig.~\ref{fig:snap_sph})
and demonstrate that the original membrane curvature $C_{\rm {ves}}=1/R_{\rm {ves}}=0.33/r_{\rm {rod}}$ changes the tubulation dynamics.
For the positive curvature $C_{\rm {rod}}r_{\rm {rod}}=4$, no tubulation is obtained for $C_{\rm {side}}r_{\rm {rod}}=-1$, $0$, and $1$.
Instead, the vesicle deforms to an elliptic disk and the rods surround the disk rim [see  Fig.~\ref{fig:snap_sph}(b)].
A discoidal bud  is often transiently formed, but the rearrangement of the rod assemblies results in disk formation even for $C_{\rm {side}}r_{\rm {rod}}=-1$ (see  Supplemental Movie~5).
Thus, outward tubulation and network formation are suppressed in small vesicles.

In contrast, for the negative curvature $C_{\rm {rod}}r_{\rm {rod}}=-4$, tubulation into the inside of the vesicle is obtained
(see Fig.~\ref{fig:snap_sph}(c) and Supplemental Movie~6).
A percolated network with $C_{\rm {side}}C_{\rm {rod}}<0$ has a significantly longer lifetime than in the flat membrane case
 [see the middle and right snapshots in Fig.~\ref{fig:snap_sph}(c)].
The coexistence of tubules and a ring is also obtained [see the left snapshot in Fig.~\ref{fig:snap_sph}(c)].
The ring stabilizes an outward bud. 
For $C_{\rm {rod}}<0$, the rods bend the membrane towards the interior (opposite to the original membrane curve), such that
the rods locally form a saddle shape in which the two principal curvatures have  opposite signs.
The network and ring structures are stabilized by the positive (opposite) $C_{\rm {side}}>0$, but not by $C_{\rm {side}}<0$.

\section*{Discussion}

We have revealed that, in addition to the spontaneous curvature along the protein rods  $C_{\rm {rod}}$,
the perpendicular spontaneous curvature  $C_{\rm {side}}$ significantly influences the protrusion of membrane tubules.
The percolated-network structure of the rod assembly has a long lifetime for $C_{\rm {side}}<0$,
because the saddle membrane shape at branches of the rod network is stabilized 
by the opposite curvature of $C_{\rm {side}}$ with respect to $C_{\rm {rod}}$.
Thus, the network formation decelerates the tubulation significantly, despite having a minor effect 
on the equilibrium property.
Both positive surface tensions and membranes originally bending in the same direction as $C_{\rm {side}}<0$
can stabilize the network structure.
Our findings provide new insights into the regulation of biomembrane shapes by curvature-inducing proteins.

Here, we employ Langevin dynamics, in which hydrodynamic interactions are neglected.
Since the static stability of the network branch is the key factor,
we do not expect the obtained $C_{\rm {side}}$ dependence to be qualitatively changed by the hydrodynamic interactions.
However, the network formation condition may be modified.
The diffusion coefficient of the proteins on the membrane depends on the protein size, and fast
 protein diffusion compared to the membrane deformation speed likely enhances the network formation.

The F-BAR domain of Pascin is considered to have a nonzero side curvature,
since it has an S-shape on the membrane surface in the addition to the curvature perpendicular to the membrane \cite{wang09}. 
Pascin induces membrane tubes in a wide range of diameter \cite{wang09,rame13}.
In the present simulations, we did not obtain such a behaviour.
It may require a larger side curvature or attractive interactions between the rods.
The rod assembly with large side curvatures is an interesting problem for further studies.

An assembly of F-BAR proteins, Cdc15, has been observed along the contractile ring of cell division \cite{lapo11}.
Their adsorption to the inner leaflet of the plasma membrane is considered to yield a ring structure similar to that shown in Fig.~\ref{fig:snap_sph}(c).
Our study suggests that the side curvature may play an important role in the formation of neck-like structures
during cell division and membrane budding in endo/exocytosis.

\section*{Methods}

We employ one of the meshless membrane models \cite{shib11},
in which a fluid membrane is represented by a self-assembled one-layer sheet of membrane particles.
A membrane particle
has an excluded volume with diameter $\sigma$ and an orientational degree of freedom.
The solvent is implicitly accounted for by an effective potential between the membrane particles.
The mechanical properties of the fluid membrane can be varied over a wide range.
The details of the meshless membrane model and protein rods are described in Ref.~\onlinecite{shib11} and Ref.~\onlinecite{nogu14}, respectively.
In this study, we employ the parameter set used in Ref. \onlinecite{nogu14} for a membrane with isotropic spontaneous curvature $C_0=0$.
The membrane has mechanical properties typical for lipid membranes:
Bending rigidity $\kappa/k_{\rm B}T=15 \pm 1$,
tensionless membrane area per particle $a_0/\sigma^2=1.2778\pm 0.0002$,
area compression modulus $K_{\rm A}\sigma^2/k_{\rm B}T=83.1 \pm 0.4$,
and edge line tension $\Gamma\sigma/k_{\rm B}T= 5.73 \pm 0.04$.

A BAR protein is modelled  as a curved rod consisting of a chain of $N_{\rm {sg}}$ membrane particles
with $r_{\rm {rod}}=10\sigma$ and $N_{\rm {sg}}=10$.
The rod has anisotropic spontaneous curvature $C_{\rm {rod}}$ along its length
and spontaneous curvature $C_{\rm {side}}$ perpendicular to its length.
When two protein rods come into contact,  $C_{\rm {side}}$ is applied between the rods.
When the rod is surrounded by membrane particles, the spontaneous curvature $C_{\rm {side}}/2$ is applied
between the rod and neighbouring membrane particles.
A molecular dynamics with a Langevin thermostat is employed \cite{shib11,nogu11}.
The simulation results are displayed with a time unit of $\tau= r_{\rm {rod}}^2/D$,
where $D$ is the diffusion coefficient of the membrane particles in the tensionless membranes.
We use total particle numbers $N=25,600$ and $9,600$ for flat membranes and vesicles, respectively. The rod density is defined as $\phi_{\rm {rod}}=N_{\rm {rod}}N_{\rm {sg}}/N$.

The mean square cluster height $\langle z_{\rm {cl}}^2\rangle$ is calculated as follows.
A rod is considered to belong to a cluster
when the distance between the centres of mass of the rod and one of the rods in the cluster is less than $r_{\rm {rod}}/2$. 
The height variance of each cluster is calculated as
$z_{\rm {i, cl}}^2 = \sum_k (z_k-z_{i, \rm {cm}})^2/N_{i,{\rm {cl}}}N_{\rm {sg}}$, 
where 
$N_{i, {\rm {cl}}}$ is the number of rods belonging to the $i$-th cluster,
 $z_{i, \rm {cm}}$ is the $z$ component of the centre of mass of the cluster,
and the summation is taken over all rod segments in the cluster.
Finally, $z_{\rm {cl}}^2$ is calculated as the average of $z_{\rm {i, cl}}^2$ for all clusters.

\section*{Acknowledgments}
This work was partially supported by a Grant-in-Aid for Scientific Research on Innovative Areas 
``Fluctuation \& Structure'' (No. 25103010) from MEXT, Japan.

\section*{Additional information}
Supplementary Information accompanies this paper at http://www.nature.com/srep

\end{document}